\documentclass[
hf,
]{ceurart}

\usepackage{listings}
\usepackage{dirtytalk}

\usepackage[ruled,vlined]{algorithm2e}

\begin{document}

\copyrightyear{2022}
\copyrightclause{Copyright for this paper by its authors.
  Use permitted under Creative Commons License Attribution 4.0
  International (CC BY 4.0).}

\conference{Woodstock'22: Symposium on the irreproducible science,
  June 07--11, 2022, Woodstock, NY}

\title{DBLP-QuAD: A Question Answering Dataset over the DBLP Scholarly Knowledge Graph}


\author[1]{Debayan Banerjee}[%
email=debayan.banerjee@uni-hamburg.de
]
\address[1]{Universität Hamburg, Hamburg, Germany}
\author[1]{Sushil Awale}[%
email=sushil.awale@studium.uni-hamburg.de,
]
\author[1]{Ricardo Usbeck}[%
email=ricardo.usbeck@uni-hamburg.de
]
\author[1]{Chris Biemann}[%
email=chris.biemann@uni-hamburg.de
]


\begin{abstract}
 In this work we create a question answering dataset over the DBLP scholarly knowledge graph (KG). DBLP is an on-line reference for bibliographic information on major computer science publications that indexes over 4.4 million publications published by more than 2.2 million authors. Our dataset consists of 10,000 question answer pairs with the corresponding SPARQL queries which can be executed over the DBLP KG to fetch the correct answer. DBLP-QuAD is the largest scholarly question answering dataset.
\end{abstract}


\maketitle

\section{Introduction}

Over the past decade, knowledge graphs (KG) such as Freebase \cite{bollacker2008freebase}, DBpedia \cite{lehmann2015dbpedia}, and Wikidata\cite{vrandevcic2014wikidata} have emerged as important repositories of general information. They store facts about the world in the linked data architecture, commonly in the format of <subject predicate object> triples. These triples can also be visualised as node-edge-node molecules of a graph structure. Much interest has been generated in finding ways to retrieve information from these KGs. Question Answering over Knowledge Graphs (KGQA) is one of the techniques used to achieve this goal. In KGQA, the focus is generally on translating a natural language question to a formal logical form. This task has, in the past, been achieved by rule-based systems \cite{asknow}. More recently, neural network and machine learning based methods have gained popularity \cite{https://doi.org/10.48550/arxiv.1907.09361}.

A scholarly KG is a specific class of KGs that contains bibliographic information. Some well known scholarly KGs are the Microsoft Academic Graph\footnote{\url{https://www.microsoft.com/en-us/research/project/microsoft-academic-graph/}}, OpenAlex\footnote{\url{http://openalex.org/}}, ORKG\footnote{\url{https://orkg.org/}} and DBLP\footnote{\url{https://dblp.org/}}. DBLP caters specifically to the bibliography of computer science, and as a result, it is smaller in size than other scholarly KGs. We decided to build our KGQA dataset over DBLP due to its focused domain and manageable size so that we could concentrate on adding complexity to the composition of the KGQA dataset itself.

Datasets are important, especially for ML-based systems, because such systems often have to be trained on a sample of data before they can be used on a similar test set. To this end, several KGQA datasets exist \cite{perevalov-etal-2022-knowledge}. However, not all datasets contain a mapping of natural language questions to the logical form (e.g. SPARQL, $\lambda$-calculus, S-expression). Some simply contain the question and the eventual answer. Such datasets can not be used to train models in the task of semantic parsing.

In this work, we present a KGQA dataset called DBLP-QuAD, which consists of 10,000 questions with corresponding SPARQL queries. The question formation process begins with human-written templates, and later, we machine-generate more questions from these templates. DBLP-QuAD consists of a variety of simple and complex questions and also tests the compositional generalisation of the models. DBLP-QuAD is the largest scholarly KGQA dataset being made available to the public\footnote{\url{https://doi.org/10.5281/zenodo.7643971}}.

\section{Related Work}

ORKG-QA benchmark \cite{jaradeh2020question} is the first scholarly KGQA dataset grounded to ORKG. The dataset was prepared using the ORKG API and focuses on the content of academic publications structured in comparison tables. The dataset is relatively small in size with only 100 question-answer pairs covering only 100 research publications.

Several other QA datasets exist, both for IR-based QA \cite{rajpurkar2018know, kwiatkowski2019natural} and KGQA \cite{trivediLCQuADCorpusComplex2017, sen2022mintaka} approaches. Several different approaches have been deployed to generate the KGQA datasets. These approaches range from manual to machine generation. However, most datasets lie in between and use a combination of manual and automated process.

A clear separation can be created between datasets that contain logical forms and those that do not. Datasets that do not require logical forms can be crowd-sourced and such datasets are generally large in size. Crowd sourcing is generally not possible for annotating logical forms because this task requires high domain expertise and it is not easy to find such experts on crowd sourcing platforms. We focus on datasets that contain logical forms.  

Free917 and QALD \cite{cai2013large,usbeck7thOpenChallenge2017} datasets were created manually by domain experts, however, their sizes are relatively small (917 and 806 respectively).

WebQuestionsSP and ComplexWebQuestions \cite{yihValueSemanticParse2016, talmorWebKnowledgebaseAnswering2018} are developed using exisiting datasets. WebQuestionsSP is a semantic parsing dataset developed by using questions from WebQuestions \cite{berantSemanticParsingFreebase2013}. \citet{yihValueSemanticParse2016} developed a dialogue-like user interface which allowed five expert human annotators to annotate the data in stages.

ComplexWebQuestions is a collection of 34,689 complex question paired with answers and SPARQL queries grounded to Freebase KG. The dataset builds on WebQuestionsSP  by sampling question-query pairs from the dataset and automatically generating questions and complex SPARQL queries with composition, conjunctions, superlatives, and comparatives functions. The machine generated questions are manually annotated to natural questions and validated by 200 AMT crowd workers.

The OVERNIGHT (ON) approach is a semantic parsing dataset generation framework introduced by \citet{wang2015building}. In this approach, the question-logical form pairs are collected with a three step process. In the first step, the logical forms are generated from a KG. Secondly, the logical forms are converted automatically into canonical questions. These canonical questions are grammatically incorrect but successfully carry the semantic meaning. Lastly, the canonical questions are converted into natural forms via crowdsourcing. Following are some of the datasets developed using this approach.

GraphQuestions \cite{suGeneratingCharacteristicrichQuestion2016} consists of 5,166 natural questions accompanied by two paraphrases of the original question, an answer, and a valid SPARQL query grounded against the Freebase KG. GraphQuestions uses a semi-automated three-step algorithm to generate the natural questions for the KG. 

LC-QuAD 1.0 \cite{trivediLCQuADCorpusComplex2017} is another semantic parsing dataset for the DBpedia KG. LC-QuAD 1.0 is relatively larger in size with 5,000 natural language English questions and corresponding SPARQL queries. The generation process starts with the set of manually created SPARQL query templates, a list of seed entities, and a whitelist of predicates. Using the list of seed entities, two-hop subgraphs from DBpedia are extracted. The SPARQL query templates consist of placeholders for both entities and predicates which are instantiated using triples from the subgraph. These SPARQL queries are then used to instantiate natural question templates which form the base for manual paraphrasing by humans. 

LC-QuAD 2.0 \cite{dubeyLCQuADLargeDataset2019} is the second iteration of LC-QuAD 1.0 with 30,000 questions, their paraphrases and their corresponding SPARQL queries compatible with both Wikidata and DBpedia KGs. Similar to LC-QuAD 1.0, in LC-QuAD 2.0 a sub-graph is generated using seed entities and a SPARQL query template is selected based on whitelist predicates. Then, the query template is instantiated using the sub-graph. Next, a template question is generated from the SPARQL query which is then verbalised and paraphrased by AMT crowd workers. LC-QuAD 2.0 has more questions and more variation compared to LC-QuAD 1.0 with paraphrases to the natural questions. 

GrailQA \cite{guThreeLevelsGeneralization2021} extends the approach in  \cite{suGeneratingCharacteristicrichQuestion2016} to generate 64,331 question-S-expression pairs grounded to the Freebase Commons KG. Here, S-expression are linearized forms of graph queries. Query templates extracted from graph queries generated from the KG are used to generate canonical logical forms grounded to compatible entities. The canonical logic forms are then validated by a graduate student if they represent plausible user query or not. Next, another graduate student annotated the validated canonical logic form with a canonical question. Finally, 6,685 Amazon Mechanical Turk workers write five natural paraphrases for each canonical question which are further validated by multiple independent crowd workers. 

KQA Pro \cite{caoKQAProDataset2022} is a large collection of 117,000 complex questions paired with
SPARQL queries for the Wikidata KG. KQA Pro dataset also follows the OVERNIGHT approach where
firstly facts from the KG are extracted. Next, canonical questions are generated with corresponding
SPARQL queries, ten answer choices and a golden answer. The canonical questions are then converted
into natural language with paraphrases using crowd sourcing. 

CFQ \cite{keysersMeasuringCompositionalGeneralization2020} (Compositional Freebase Questions) is a semantic parsing dataset developed completely using synthetic generation approaches that consists of simple natural language questions with corresponding SPARQL query against the Freebase KG. CFQ contains 239,357 English questions which are generated using hand-crafted grammar and inference rules with a corresponding logical form. Next, resolution rules are used to map the logical forms to SPARQL queries. The CFQ dataset was specifically designed to measure compositional generalization.

In this work, we loosely follow the OVERNIGHT approach to create a large scholarly KGQA dataset for the DBLP KG.

\section{DBLP KG}

\begin{figure*}[ht]
\centering
    \includegraphics[width=0.8\linewidth]{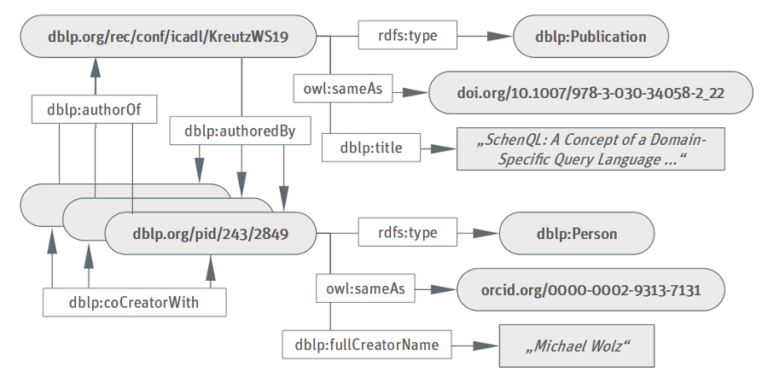}
\caption{Example of entries in the DBLP KG with its schema}
\label{dblp1}
\end{figure*}

DBLP, which used to stand for Data Bases and Logic Programming\footnote{\url{https://en.wikipedia.org/wiki/DBLP}}, was created in 1993 by Michael Ley at the University of Trier, Germany \cite{leyDBLPComputerScience2002}. The service was originally designed as a bibliographic database for research papers and proceedings from the fields of database systems and logic programming. Over time, the service has grown in size and scope, and today includes bibliographic information on a wide range of topics within the field of computer science. The DBLP RDF data models a person-publication graph shown in Figure \ref{dblp1}.


The DBLP KG contains two main entities: \textit{Person} and \textit{Publication}, where as other metadata such as journal and conferences, affiliation of authors are currently only string literals. Henceforth, we use the term \textit{person} and \textit{creator} interchangeably. At the time of its release, the RDF dump consisted of 2,941,316 person entities, 6,010,605 publication entities, and 252,573,199 RDF triples. DBLP currently does not provide a SPARQL endpoint but the RDF dump can be downloaded and a local SPARQL endpoint such as Virtuoso Server can be setup to run a SPARQL query against the DBLP KG.

The live RDF data model on the DBLP website follows the schema shown in Figure \ref{dblp1}. However, the RDF snapshots available for download have the \textit{coCreatorWith} and \textit{authorOf} predicates missing. Although these predicates are missing, the \textit{authoredBy} predicate can be used to derive the missing relations. DBLP-QuAD is based on the DBLP KG schema of the downloadable RDF graph.

\section{Dataset Generation Framework}

\begin{figure*}[h]
\centering
    \includegraphics[width=0.89\linewidth]{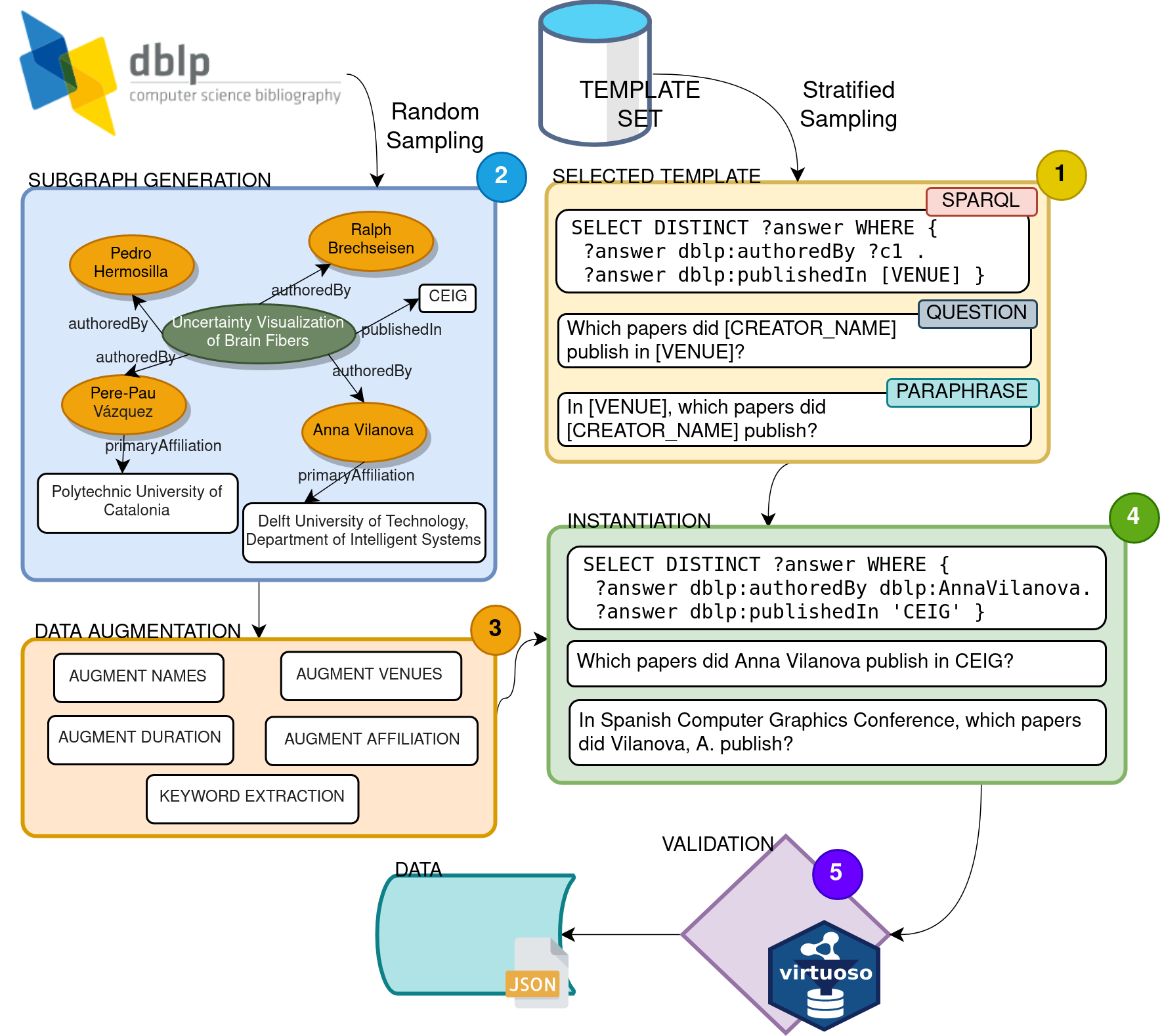}
\caption{\textbf{Motivating Example}. The generation process starts with (1) selection of a template tuple followed by (2) subgraph generation. Then, literals in subgraph are (3) augmented before being used to (4) instantiate the selected template tuple. The generated data is (5) filtered based on if they produce answers or not.}
\label{datagen}
\end{figure*}

In this work, the aim is to generate a large variety of scholarly questions and corresponding SPARQL query pairs for the DBLP KG. Initially, a small set of templates $T$ containing a SPARQL query template $s_t$ and a few semantically equivalent natural language question templates $Q_t$ are created. The questions and query templates are created such that they cover a wide range of scholarly metadata user information need while also being answerable using a SPARQL query against the DBLP KG. Next, we synthetically generate a large set of question-query pairs $(q_i, s_i)$ suitable for training a neural network semantic parser.

The core methodology of the dataset generation framework encompasses instantiating the templates using literals of subgraphs sampled from the KG. Moreover, to capture different representations of the literal values from a human perspective, we randomly mix in different augmentations of these textual representations. The dataset generation workflow is shown in Figure \ref{datagen}.

\subsection{Templates}

The first step in the dataset generation process starts with the creation of a template set. After carefully analyzing the ontology of the DBLP KG, we manually wrote 98 pairs of valid SPARQL query templates and a set of semantically equivalent natural language question templates. The template set was written by one author and verified for correctness by another author. The query and question templates consist of placeholder markers instead of URIs, entity surface forms or literals. For example, in Figure \ref{datagen} (Section $1$), the SPARQL query template includes the placeholders $?c1$ and $[VENUE]$ for DBLP person URI and venue literal respectively. Similarly, the question templates include placeholders $[CREATOR\_NAME]$ and $[VENUE]$ for creator name and venue literal respectively. The template set covers the two entities creator and publication, and additionally the foreign entity bibtex type. Additionally, they also cover the $11$ different predicates of DBLP KG. 

The template set consists of template tuples. A template tuple $t = (s_t, Q_t, E_t, P_t)$ is composed of a SPARQL query template $s_t$, a set of semantically equivalent natural language question templates $Q_t$, a set of entity placeholders $E_t$ and a set of predicates $P_t$ used in $s_t$. We also add a boolean indicating whether the query template is temporal or not and another boolean indicating whether to use or not use the template while generating $train$ dataset. Each template tuple contains between four and seven paraphrased question templates offering wide linguistic diversity. While most of the question templates use the \textit{"Wh-"} question keyword, we also include instruction-style paraphrases.

We group the template tuples as creator-focused or publication-focused $\epsilon$ and further group them by query types $\delta$. We have $10$ different query types and they include Single Fact, Multiple Facts, Boolean, Negation, Double Negation, Double Intent, Union, Count, Superlative/Comparative, and Disambiguation. The question types are discussed in Section \ref{questions} with examples. The distribution of templates per entity and query type is shown in Table \ref{tab:template_dist}. During dataset generation, for each data instance we sample a template tuple from the template set using stratified sampling maintaining equal distribution of entity types and query types.

\begin{table*}[ht]
    \centering
    \begin{tabular}{c|c|c|c}
        \hline
         Query Type & Creator-focused & Publication-focused & Total \\
         \hline
         Single Fact & 5 & 5 & 10 \\
         Multiple Facts & 7 & 7 & 14 \\
         Boolean & 6 & 6 & 12 \\
         Negation & 4 & 4 & 8 \\
         Double Negation & 4 & 4 & 8 \\
         Double Intent & 5 & 4 & 9 \\
         Union & 4 & 4 & 8 \\
         Count & 6 & 5 & 11 \\
         Superlative/Comparative & 6 & 6 & 12 \\
         Disambiguation & 3 & 3 & 6\\
         \hline
         Total & 50 & 48 & 98
    \end{tabular}
    \caption{Total number of template tuples per query type grouped by entity type}
    \label{tab:template_dist}
\end{table*} 

\subsection{Subgraph generation}

The second part of the dataset generation framework is subgraph generation. Given a graph $G=(V,E)$ where $V$ are the vertices, and $E$ are edges, we draw a subgraph $g=(v,e)$ where $v \subset V$, $e \subset E$. For the DBLP KG, $V$ are the creator and publication entity URIs or literals, and the $E$ are the predicates of the entities.

The subgraph generation process starts with random sampling of a publication entity $v_i$ from the DBLP KG. We only draw from the set of publication entities as the RDF snapshot available for download has $authorOf$ and $coCreatorWith$ predicates missing for creator entity. As such, a subgraph centered on a creator entity would not have end vertices that can be expanded further. With the sampled publication entity $v_i$, we iterate through all the predicates $e$ to extract creator entities $v'$ as well as the literal values. We further, expand the creator entities and extract their literal values to form a two-hop subgraph $g=(v, e)$ as shown in Figure \ref{datagen} (Section $2$).

\subsection{Template Instantiation}

Using the generated subgraph and the sampled template tuple, the template tuple is instantiated with entity URIs and literal values from the subgraph. In the instantiation process, a placeholder marker in a string is replaced by the corresponding text representation.

For the SPARQL query template $s_t$, we instantiate the creator/publication placeholder markers with DBLP creator/publication entity URIs or literal values for affiliation and conference or journals to create a valid SPARQL query $s$ that returns answers when run against the DBLP KG SPARQL endpoint.

In case of natural language question templates, we randomly sample two from the set of question templates $q_t^1, q_t^2 \in Q_T$, and instantiate each using only the literal values from the subgraph to form one main natural language question $q^1$ and one natural language question paraphrase $q^2$. In natural language, humans can write the literal strings in various forms. Hence to introduce this linguistic variation, we randomly mix in alternate string representations of these literal values in both natural language questions. The data augmentation process allows us to add heuristically manipulated alternate literal representations to the natural questions. A example of an instantiated template is shown in Figure \ref{datagen} (Section $3$).

\subsection{Data Augmentation}

For the template instantiation process, we perform simple string manipulations to generate alternate literal representations. Then, we randomly select between the original literal representation and the alternate representation to instantiate the natural language questions. For each literal type, we apply different string manipulation techniques which we describe below.

\textbf{Names}: For names we generate four different alternatives involving switching parts of names or keeping only initials of the names. Consider the name \textit{John William Smith} for which we produce \textit{Smith, John William}, \textit{J. William Smith}, \textit{John W. Smith}, and \textit{Smith, J. William}.

\textbf{Venues}: Venues can be represented using either its short form or its full form. For example, \textit{ECIR} or \textit{European Conference on Information Retrieval}. In DBLP venues are stored in its short form. We use a selected list of conference and journals\footnote{\url{http://portal.core.edu.au/conf-ranks/?search=&by=all&source=CORE2021&sort=atitle&page=1}} containing the short form and its equivalent full form to get the full venue names.

\textbf{Duration}: About 20\% of the templates contain temporal queries, and some of them require dummy numbers to represent duration. For example, the question \textit{"In the last five years, which papers did Mante S. Nieuwland publish?"} uses the dummy value \textit{five}. We randomly select between the numerical representation and the textual representation for the dummy duration value.

\textbf{Affiliation}: In natural language questions, only the institution name is widely used to refer to the affiliation of an author. However, the DBLP KG uses the full address of an institution including city and country name. Hence, using RegeEx we extract the institution names and randomly select between the institution name and the full institution address in the instantiation process.

\textbf{Keywords}: For disambiguation queries, we do not use the full title of a publication but rather a part of it by extracting keywords. For this purpose, we use SpaCy's Matcher API\footnote{\url{https://spacy.io/api/matcher/}} to extract noun phrases from the title.

\begin{algorithm}
    \caption{Dataset Generation Process}
    \label{algo:main}
    
    \SetKwInOut{Input}{inputs}
    \SetKwInOut{Output}{output}
    \SetKwProg{GenerateDataset}{GenerateDataset}{}{}

    \GenerateDataset{$(T, x, N, G)$}{
        \Input{template set $T$; dataset set to generate $x$; size of dataset to generate $N$; KG to sample subgraphs from $G$;}
        \Output{dataset $D$\;}
        $D \gets \emptyset$\;
        $n \gets (N / |\epsilon|) / |\delta|$\;
        \ForEach{$e \in \epsilon$}{
            \ForEach{$s \in \delta$}{
                $i \gets 0$\;
                $T_{es} \gets T[e][s]$\;
                \If{$x == train$}{
                    $T_{es} \gets Filter(T_{es}, \space test\_only \space == \space True)$
                }
                \While{$i < n$}{
                    $g_1, g_2 \gets SampleSubgraph(G, \space 2)$\; 
                    $t_i \gets random.sample(T_{es})$\; 
                    $d_i \gets Instantiate(t_i, g_1, g_2, x)$\;
                    $answer \gets Query(d_i)$\;
                    \If{$answer$}{ 
                        $D \gets d_i$\;
                        $i \gets i + 1$\;
                    }
                }
            }
        }
        \KwRet{D}
    }
\end{algorithm}

\subsection{Dataset Generation}

For each data instance $d_i$, we sample $2$ subgraphs (\textit{SampleSubgraph(G,2)}) and instantiate a template tuple $t_i$ (\textit{{Instantiate($t_i$, $g_1$, $g_2$, x)}}). We sample $2$ subgraphs as some template tuples require to be instantiated with two publication titles. Each data instance $d_i = (s_i,q^1_i,q^2_i,E_i,P_i,y,z)$ comprises of a valid SPARQL query $s_i$, one main natural language question $q^1_i$, one semantically equivalent paraphrase of the main question $q^2_i$, a list of entities $E_i$ used in $s_i$, a list of predicates $P_i$ used in $s_i$, a Boolean indicating whether the SPARQL query is temporal or not $y$, and another Boolean informing whether the SPARQL query is found only in $valid$ and $test$ sets $z$. We generate an equal number $n$ of questions for each entity group $\epsilon$ equally divided for each query type $\delta$.

To foster a focus on generalization ability, we manually marked $20$ template tuples to withhold during generation of the $train$ set. However, we use all the template tuples in the generation of $valid$ and $test$ sets. Furthermore, we also withhold $2$ question templates when generating $train$ questions but use all question templates when generating $valid$ and $test$ sets. This controlled generation process allows us to withhold some entity classes, predicates and paraphrases from $train$ set. Our aim with this control is to create a scholarly KGQA dataset that facilitates development of KGQA models that adhere to \textit{i.i.d}, \textit{compositional}, and \textit{zero-shot} \cite{guThreeLevelsGeneralization2021} generalization.

Further, we validate each data instance $d_i$ by running the SPARQL query $s_i$ against the DBLP KG via a Virtuoso SPARQL endpoint\footnote{\url{https://docs.openlinksw.com/virtuoso/whatisvirtuoso/}}. We filter out data instances for which the SPARQL query is invalid or generates a blank response. A SPARQL query may generate a blank response if the generated subgraphs have missing literal values. In the DBLP KG, some of the entities have missing literals for predicates such as \textit{primaryAffiliation}, \textit{orcid}, \textit{wikidata}, and so on. Additionally, we also store the answers produced by the SPARQL query against the DBLP KG formatted according to \textit{\url{https://www.w3.org/TR/sparql11-results-json/}}. 
The dataset generation process is summarized in Algorithm \ref{algo:main}.

\subsection{Types of Questions}
\label{questions}

The dataset is composed of the following question types. The examples shown here are hand-picked from the dataset.

\begin{itemize}
    \item \textbf{Single fact}: These questions can be answered using a single fact. For example, \say{What year was \say{SIRA: SNR-Aware Intra-Frame Rate Adaptation} published?}
    
    \item \textbf{Multiple facts}: These questions require connecting two or more facts to answer. For example, \say{In SIGCSE, which paper written by Darina Dicheva with Dichev, Christo was published?}
    
    \item \textbf{Boolean}: These questions answer where a given fact is true or false. We can also add negation keywords to negate the questions. For example, \say{Does Szeider, Stefan have an ORCID?}
    
    \item \textbf{Negation}: These questions require to negate the answer to the Boolean questions. For example, \say{Did M. Hachani not publish in ICCP?}
    
    \item \textbf{Double negation}: These questions require to negate the Boolean question answers twice which results. For example, \say{Wasn't the paper \say{Multi-Task Feature Selection on Multiple Networks via Maximum Flows} not published in 2014?}
    
    \item \textbf{Count}: These questions pertain to the count of occurrence of facts. For example, \say{Count the authors of \say{Optimal Symmetry Breaking for Graph Problems} who have Carnegie Mellon University as their primary affiliation.}
    
    \item \textbf{Superlative/Comparative}: Superlative questions ask about the maximum and minimum for a subject and comparative questions compare values between two subjects. We group both types under one group. For example, \say{Who has published the most papers among the authors of \say{k-Pareto optimality for many-objective genetic optimization}?}
    
    \item \textbf{Union} questions cover a single intent but for multiple subjects at the same time. For example, \say{List all the papers that Pitas, Konstantinos published in ICML and ISCAS.}
    
    \item \textbf{Double intent} questions poses two user intentions, usually about the same subject. For example, \say{In which venue was the paper \say{Interactive Knowledge Distillation for image classification} published and when?}
    
    \item \textbf{Disambiguation} questions requires identifying the correct subject in the question. For example, \say{Which author with the name Li published the paper about Buck power converters?}
\end{itemize}

\section{Dataset Statistics}

DBLP-QuAD consists of 10,000 unique question-query pairs grouped into \textit{train}, \textit{valid} and \textit{test} sets with a ratio of \textit{7:1:2}. The dataset covers 13,348 creators and publications, and 11 predicates of the DBLP KG. For each query type in Table \ref{tab:template_dist}, the dataset includes 1,000 question-query pairs each of which is equally divided as creator-focused or publication-focused. Additionally, among the questions in DBLP-QuAD, 2,350 are temporal questions.

\textbf{Linguistic Diversity.} In DBLP-QuAD, a natural language question has an average word length of 17.32 words and an average character length of 114.1 characters. Similarly, a SPARQL query has an average vocab length of 12.65 and an average character length of 249.48 characters. Between the natural language question paraphrases, the average Jaccard similarity for unigram and bigram are $0.62$ and $0.47$ (with standard deviations of $0.22$ and $0.24$) respectively. The average Levenshtein edit distance between them is $32.99$ (with standard deviation of $23.12$). We believe the metrics signify a decent level of linguistic diversity. 

\textbf{Entity Linking.} DBLP-QuAD also presents challenging entity linking with data augmentation performed on literals during the generation process. The augmented literals present more realistic and natural representation of the entity surface forms and literals compared to the entries in the KG.  

\textbf{Generalization.} In the \textit{valid} set 18.9\% and in the \textit{test} set 19.3\% of instances were generated using the withheld templates. Hence, these SPARQL query templates and natural language question templates are unique to the \textit{valid} and \textit{test} sets. Table \ref{tab:generalization} shows the percent of questions with different levels of generalization in the \textit{valid} and \textit{test} sets of the dataset.

\begin{table*}[ht]
    \vspace{-3mm}
    \centering
    \begin{tabular}{c|c|c|c}
         \hline
         Dataset & I.I.D & Compositional & Zero-shot  \\
         \hline
         Valid & 82.8\% & 13.6\% & 3.6\% \\
         Test & 81.2\% & 15.1\% & 3.8\% \\
         \hline
    \end{tabular}
    \caption{Percent of questions with different levels of generalization in the \textit{valid} and \textit{test} sets of DBLP-QuAD}
    \label{tab:generalization}
    \vspace{-3mm}
\end{table*}

\section{Semantic Parsing Baseline}

To lay the foundation for future work on DBLP-QuAD, we also release baselines using the recent work by \citet{banerjeeModernBaselinesSPARQL2022}, where a pre-trained T5 model is fine-tuned \cite{raffelExploringLimitsTransfer} on the LC-QuAD 2.0 dataset.

\begin{figure*}[ht]
    \centering
    \includegraphics[width=0.80\linewidth]{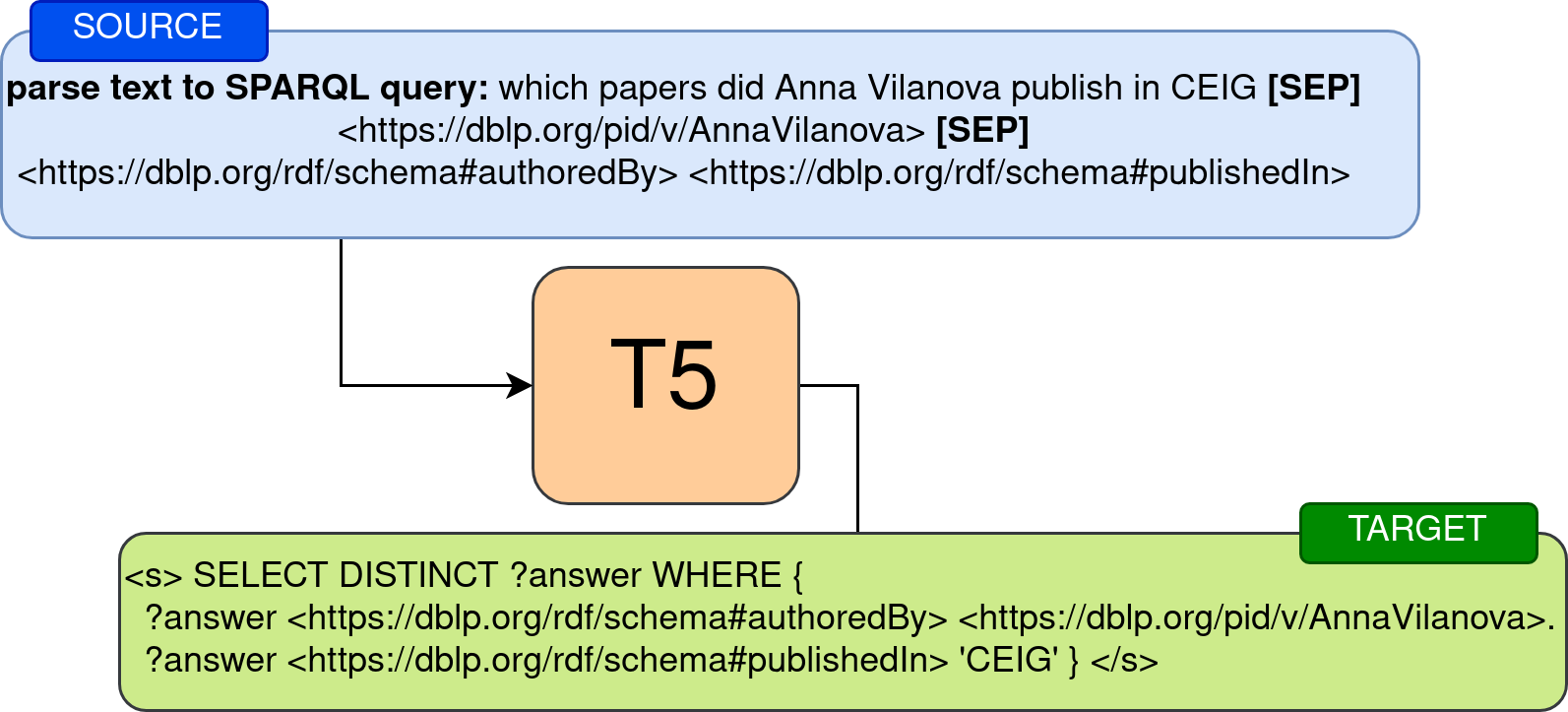}
    \caption{Representation of source and target text used to fine-tune the T5 model}
    \label{fig:t5}
\end{figure*}

Following \citet{banerjeeModernBaselinesSPARQL2022}, we assume the entities and the relations are linked, and only focus on query building. We formulate the source as shown in Figure \ref{fig:t5}, where for each natural language question a prefix \say{\textbf{parse text to SPARQL query:}} is added. The source string is further concatenated with entity URIs and relation schema URIs separated by a special token $[SEP]$. The target text is the corresponding SPARQL query which is padded with the tokens $<s></s>$. We also make use of the sentinel tokens provided by T5 to represent the DBLP prefixes e.g. \textit{<extra\_id\_1>} denotes the prefix \textit{https://dblp.org/pid/}, SPARQL vocabulary and symbols. This step helps the \textit{T5-tokenizer} to correctly fragment the target text during inference.

We fine-tune \textit{T5-Base} and \textit{T5-Small} on DBLP-QuAD train set with a learning rate of \textit{1e-4} for $5$ epochs with an input as well as output text length of $512$ and batch size of $4$. 

\subsection{Experiment Results}

We report the performance of the baseline model on the DBLP-QuAD test set. Firstly, we report on the exact-match between the gold and the generated SPARQL query. For the exact-match accuracy we compare the generated and the gold query token by token after removing whitespaces. Next, for each SPARQL query on the test set, we run both the gold and and the query generated by the T5 baseline models using Virtuoso SPARQL endpoint to fetch answers from the DBLP KG. Based on the answers collected, we report on the F1 score. The results are reported on Table \ref{tab:results}.

\begin{table*}[ht]
    \centering
    \begin{tabular}{c|c|c}
         \hline
         Evaluation metrics & T5-Small & T5-Base \\
         \hline
         Exact-match Accuracy & 0.638 & 0.813 \\
         F1 Score & 0.721 & 0.868 \\
         \hline
    \end{tabular}
    \caption{Evaluation results of fine-tuned T5 to DBLP-QuAD}
    \label{tab:results}
\end{table*}

\vspace{-5mm}
\section{Limitations}
\vspace{-3mm}

One of the drawbacks of our dataset generation framework is that natural questions are synthetically generated. (CFQ \cite{keysersMeasuringCompositionalGeneralization2020} has a similar limitation.) Although the question templates were human-written, only two people (authors of the paper) worked on the creation of the question templates and was not crowd sourced from a group of researchers. Additionally, the questions are generated by drawing data from a KG. Hence, the questions may not perfectly reflect the distribution of user information need. However, the machine-generation process allows for programmatic configuration of the questions, setting question characteristics, and controlling dataset size. We utilize the advantage by programmatically augmenting text representations and generating a large scholarly KGQA with complex SPARQL queries. 

Second, in generating \textit{valid} and \textit{test} sets, we utilize additional 19 template tuples which account for about 20\% of the template set. Therefore, the syntactic structure for 80\% of the generated data in \textit{valid} and \textit{test} would already be seen in the train set resulting in test leakage. However, to limit the leakage on 80\% of the data, we withhold $2$ question templates in generating the $train$ set. Moreover, the data augmentation steps carried out would also add challenges in the $valid$ and $test$ sets.

Another shortcoming of DBLP-QuAD is that the paper titles do not perfectly reflect user behavior. When a user asks a question, they do not type in the full paper title and also some papers are popularly known by a different short name. For example, the papers \say{Language Models are Few-shot Learners} and \say{BERT: Pre-training of Deep Bidirectional Transformers for Language Understanding} are also known as \say{GPT-3} and \say{BERT} respectively. This is a challenging entity linking problem which requires further investigation. Despite the shortcomings, we feel the large scholarly KGQA dataset would ignite more research interest in scholarly KGQA.

\vspace{-3mm}
\section{Conclusion}
\vspace{-3mm}

In this work, we presented a new KGQA dataset called DBLP-QuAD. The dataset is the largest scholarly KGQA dataset with corresponding SPARQL queries. The dataset contains a wide variety of questions and query types and we present the data generation framework and baseline results. We hope this dataset proves to be a valuable resource for the community.

As future work, we would like to build a robust question answering system for scholarly data using this dataset.

\vspace{-3mm}
\section{Acknowledgements}
\vspace{-3mm}

This research was supported by grants from NVIDIA and utilized NVIDIA 2 x RTX A5000 24GB. Furthermore, we acknowledge the financial support from the Federal Ministry for Economic Affairs and Energy of Germany in the project CoyPu (project number 01MK21007[G]) and the German Research Foundation in the project NFDI4DS (project number 460234259). This research is additonally funded by the ``Idea and Venture Fund`` research grant by Universit\"at Hamburg, which is part of the Excellence Strategy of the Federal and State Governments.

\bibliography{sample-ceur,sushilrefs}

\begin{thebibliography}{25}
\expandafter\ifx\csname natexlab\endcsname\relax\def\natexlab#1{#1}\fi
\providecommand{\url}[1]{\texttt{#1}}
\providecommand{\href}[2]{#2}
\providecommand{\path}[1]{#1}
\providecommand{\DOIprefix}{doi:}
\providecommand{\ArXivprefix}{arXiv:}
\providecommand{\URLprefix}{URL: }
\providecommand{\Pubmedprefix}{pmid:}
\providecommand{\doi}[1]{\href{http://dx.doi.org/#1}{\path{#1}}}
\providecommand{\Pubmed}[1]{\href{pmid:#1}{\path{#1}}}
\providecommand{\bibinfo}[2]{#2}
\ifx\xfnm\relax \def\xfnm[#1]{\unskip,\space#1}\fi
\bibitem[{Bollacker et~al.(2008)Bollacker, Evans, Paritosh, Sturge, and
  Taylor}]{bollacker2008freebase}
\bibinfo{author}{K.~Bollacker}, \bibinfo{author}{C.~Evans},
  \bibinfo{author}{P.~Paritosh}, \bibinfo{author}{T.~Sturge},
  \bibinfo{author}{J.~Taylor},
\newblock \bibinfo{title}{{Freebase: A Collaboratively Created Graph Database
  for Structuring Human Knowledge}},
\newblock in: \bibinfo{booktitle}{Proceedings of the 2008 ACM SIGMOD
  international conference on Management of data}, \bibinfo{organization}{AcM},
  \bibinfo{year}{2008}, pp. \bibinfo{pages}{1247--1250}.
\bibitem[{Lehmann et~al.(2015)Lehmann, Isele, Jakob, Jentzsch, Kontokostas,
  Mendes, Hellmann, Morsey, Van~Kleef, Auer et~al.}]{lehmann2015dbpedia}
\bibinfo{author}{J.~Lehmann}, \bibinfo{author}{R.~Isele},
  \bibinfo{author}{M.~Jakob}, \bibinfo{author}{A.~Jentzsch},
  \bibinfo{author}{D.~Kontokostas}, \bibinfo{author}{P.~N. Mendes},
  \bibinfo{author}{S.~Hellmann}, \bibinfo{author}{M.~Morsey},
  \bibinfo{author}{P.~Van~Kleef}, \bibinfo{author}{S.~Auer}, et~al.,
\newblock \bibinfo{title}{{DBpedia -- A Large-Scale, Multilingual Knowledge
  Base Extracted from Wikipedia}},
\newblock \bibinfo{journal}{Semantic Web}  (\bibinfo{year}{2015}).
\bibitem[{Vrande{\v{c}i{\'c}, Denny and Kr{\"o}tzsch,
  Markus}(2014)}]{vrandevcic2014wikidata}
\bibinfo{author}{Vrande{\v{c}i{\'c}, Denny and Kr{\"o}tzsch, Markus}},
\newblock \bibinfo{title}{{Wikidata: A Free Collaborative Knowledge Base}},
\newblock \bibinfo{journal}{Communications of the ACM}  (\bibinfo{year}{2014}).
\bibitem[{Dubey et~al.(2016)Dubey, Dasgupta, Sharma, H{\"o}ffner, and
  Lehmann}]{asknow}
\bibinfo{author}{M.~Dubey}, \bibinfo{author}{S.~Dasgupta},
  \bibinfo{author}{A.~Sharma}, \bibinfo{author}{K.~H{\"o}ffner},
  \bibinfo{author}{J.~Lehmann},
\newblock \bibinfo{title}{{AskNow: A Framework for Natural Language Query
  Formalization in SPARQL}},
\newblock in: \bibinfo{editor}{H.~Sack}, \bibinfo{editor}{E.~Blomqvist},
  \bibinfo{editor}{M.~d'Aquin}, \bibinfo{editor}{C.~Ghidini},
  \bibinfo{editor}{S.~P. Ponzetto}, \bibinfo{editor}{C.~Lange} (Eds.),
  \bibinfo{booktitle}{The Semantic Web. Latest Advances and New Domains},
  \bibinfo{publisher}{Springer International Publishing},
  \bibinfo{address}{Cham}, \bibinfo{year}{2016}, pp. \bibinfo{pages}{300--316}.
\bibitem[{Chakraborty et~al.(2019)Chakraborty, Lukovnikov, Maheshwari, Trivedi,
  Lehmann, and Fischer}]{https://doi.org/10.48550/arxiv.1907.09361}
\bibinfo{author}{N.~Chakraborty}, \bibinfo{author}{D.~Lukovnikov},
  \bibinfo{author}{G.~Maheshwari}, \bibinfo{author}{P.~Trivedi},
  \bibinfo{author}{J.~Lehmann}, \bibinfo{author}{A.~Fischer},
  \bibinfo{title}{{Introduction to Neural Network based Approaches for Question
  Answering over Knowledge Graphs}}, \bibinfo{year}{2019}. \URLprefix
  \url{https://arxiv.org/abs/1907.09361}.
  \DOIprefix\doi{10.48550/ARXIV.1907.09361}.
\bibitem[{Perevalov et~al.(2022)Perevalov, Yan, Kovriguina, Jiang, Both, and
  Usbeck}]{perevalov-etal-2022-knowledge}
\bibinfo{author}{A.~Perevalov}, \bibinfo{author}{X.~Yan},
  \bibinfo{author}{L.~Kovriguina}, \bibinfo{author}{L.~Jiang},
  \bibinfo{author}{A.~Both}, \bibinfo{author}{R.~Usbeck},
\newblock \bibinfo{title}{{Knowledge Graph Question Answering Leaderboard: A
  Community Resource to Prevent a Replication Crisis}},
\newblock in: \bibinfo{booktitle}{Proceedings of the Thirteenth Language
  Resources and Evaluation Conference}, \bibinfo{publisher}{European Language
  Resources Association}, \bibinfo{address}{Marseille, France},
  \bibinfo{year}{2022}, pp. \bibinfo{pages}{2998--3007}. \URLprefix
  \url{https://aclanthology.org/2022.lrec-1.321}.
\bibitem[{Jaradeh et~al.(2020)Jaradeh, Stocker, and Auer}]{jaradeh2020question}
\bibinfo{author}{M.~Y. Jaradeh}, \bibinfo{author}{M.~Stocker},
  \bibinfo{author}{S.~Auer},
\newblock \bibinfo{title}{Question answering on scholarly knowledge graphs},
\newblock in: \bibinfo{booktitle}{International Conference on Theory and
  Practice of Digital Libraries}, \bibinfo{organization}{Springer},
  \bibinfo{year}{2020}, pp. \bibinfo{pages}{19--32}.
\bibitem[{Rajpurkar et~al.(2018)Rajpurkar, Jia, and Liang}]{rajpurkar2018know}
\bibinfo{author}{P.~Rajpurkar}, \bibinfo{author}{R.~Jia},
  \bibinfo{author}{P.~Liang},
\newblock \bibinfo{title}{{Know what you don't know: Unanswerable questions for
  SQuAD}},
\newblock \bibinfo{journal}{arXiv preprint arXiv:1806.03822}
  (\bibinfo{year}{2018}).
\bibitem[{Kwiatkowski et~al.(2019)Kwiatkowski, Palomaki, Redfield, Collins,
  Parikh, Alberti, Epstein, Polosukhin, Devlin, Lee
  et~al.}]{kwiatkowski2019natural}
\bibinfo{author}{T.~Kwiatkowski}, \bibinfo{author}{J.~Palomaki},
  \bibinfo{author}{O.~Redfield}, \bibinfo{author}{M.~Collins},
  \bibinfo{author}{A.~Parikh}, \bibinfo{author}{C.~Alberti},
  \bibinfo{author}{D.~Epstein}, \bibinfo{author}{I.~Polosukhin},
  \bibinfo{author}{J.~Devlin}, \bibinfo{author}{K.~Lee}, et~al.,
\newblock \bibinfo{title}{{Natural questions: a benchmark for question
  answering research}},
\newblock \bibinfo{journal}{Transactions of the Association for Computational
  Linguistics} \bibinfo{volume}{7} (\bibinfo{year}{2019})
  \bibinfo{pages}{453--466}.
\bibitem[{Trivedi et~al.(2017)Trivedi, Maheshwari, Dubey, and
  Lehmann}]{trivediLCQuADCorpusComplex2017}
\bibinfo{author}{P.~Trivedi}, \bibinfo{author}{G.~Maheshwari},
  \bibinfo{author}{M.~Dubey}, \bibinfo{author}{J.~Lehmann},
\newblock \bibinfo{title}{{{LC-QuAD}}: {{A Corpus}} for {{Complex Question
  Answering}} over {{Knowledge Graphs}}},
\newblock in: \bibinfo{editor}{C.~{d'Amato}}, \bibinfo{editor}{M.~Fernandez},
  \bibinfo{editor}{V.~Tamma}, \bibinfo{editor}{F.~Lecue},
  \bibinfo{editor}{P.~{Cudr{\'e}-Mauroux}}, \bibinfo{editor}{J.~Sequeda},
  \bibinfo{editor}{C.~Lange}, \bibinfo{editor}{J.~Heflin} (Eds.),
  \bibinfo{booktitle}{The {{Semantic Web}} \textendash{} {{ISWC}} 2017}, volume
  \bibinfo{volume}{10588}, \bibinfo{publisher}{{Springer International
  Publishing}}, \bibinfo{address}{{Cham}}, \bibinfo{year}{2017}, pp.
  \bibinfo{pages}{210--218}. \DOIprefix\doi{10.1007/978-3-319-68204-4_22}.
\bibitem[{Sen et~al.(2022)Sen, Aji, and Saffari}]{sen2022mintaka}
\bibinfo{author}{P.~Sen}, \bibinfo{author}{A.~F. Aji},
  \bibinfo{author}{A.~Saffari},
\newblock \bibinfo{title}{{Mintaka: A Complex, Natural, and Multilingual
  Dataset for End-to-End Question Answering}},
\newblock \bibinfo{journal}{arXiv preprint arXiv:2210.01613}
  (\bibinfo{year}{2022}).
\bibitem[{Cai and Yates(2013)}]{cai2013large}
\bibinfo{author}{Q.~Cai}, \bibinfo{author}{A.~Yates},
\newblock \bibinfo{title}{{Large-scale semantic parsing via schema matching and
  lexicon extension}},
\newblock in: \bibinfo{booktitle}{Proceedings of the 51st Annual Meeting of the
  Association for Computational Linguistics (Volume 1: Long Papers)},
  \bibinfo{year}{2013}, pp. \bibinfo{pages}{423--433}.
\bibitem[{Usbeck et~al.(2017)Usbeck, Ngomo, Haarmann, Krithara, R{\"o}der, and
  Napolitano}]{usbeck7thOpenChallenge2017}
\bibinfo{author}{R.~Usbeck}, \bibinfo{author}{A.-C.~N. Ngomo},
  \bibinfo{author}{B.~Haarmann}, \bibinfo{author}{A.~Krithara},
  \bibinfo{author}{M.~R{\"o}der}, \bibinfo{author}{G.~Napolitano},
\newblock \bibinfo{title}{7th {{Open Challenge}} on {{Question Answering}} over
  {{Linked Data}} ({{QALD-7}})},
\newblock in: \bibinfo{editor}{M.~Dragoni}, \bibinfo{editor}{M.~Solanki},
  \bibinfo{editor}{E.~Blomqvist} (Eds.), \bibinfo{booktitle}{Semantic {{Web
  Challenges}}}, volume \bibinfo{volume}{769}, \bibinfo{publisher}{{Springer
  International Publishing}}, \bibinfo{address}{{Cham}}, \bibinfo{year}{2017},
  pp. \bibinfo{pages}{59--69}. \DOIprefix\doi{10.1007/978-3-319-69146-6_6}.
\bibitem[{Yih et~al.(2016)Yih, Richardson, Meek, Chang, and
  Suh}]{yihValueSemanticParse2016}
\bibinfo{author}{W.-t. Yih}, \bibinfo{author}{M.~Richardson},
  \bibinfo{author}{C.~Meek}, \bibinfo{author}{M.-W. Chang},
  \bibinfo{author}{J.~Suh},
\newblock \bibinfo{title}{The {{Value}} of {{Semantic Parse Labeling}} for
  {{Knowledge Base Question Answering}}},
\newblock in: \bibinfo{booktitle}{Proceedings of the 54th {{Annual Meeting}} of
  the {{Association}} for {{Computational Linguistics}} ({{Volume}} 2: {{Short
  Papers}})}, \bibinfo{publisher}{{Association for Computational Linguistics}},
  \bibinfo{address}{{Berlin, Germany}}, \bibinfo{year}{2016}, pp.
  \bibinfo{pages}{201--206}. \DOIprefix\doi{10.18653/v1/P16-2033}.
\bibitem[{Talmor and Berant(2018)}]{talmorWebKnowledgebaseAnswering2018}
\bibinfo{author}{A.~Talmor}, \bibinfo{author}{J.~Berant}, \bibinfo{title}{The
  {{Web}} as a {{Knowledge-base}} for {{Answering Complex Questions}}},
  \bibinfo{year}{2018}. \href{http://arxiv.org/abs/1803.06643}{{\tt
  arXiv:1803.06643}}.
\bibitem[{Berant et~al.(2013)Berant, Chou, Frostig, and
  Liang}]{berantSemanticParsingFreebase2013}
\bibinfo{author}{J.~Berant}, \bibinfo{author}{A.~Chou},
  \bibinfo{author}{R.~Frostig}, \bibinfo{author}{P.~Liang},
\newblock \bibinfo{title}{Semantic {{Parsing}} on {{Freebase}} from
  {{Question-Answer Pairs}}},
\newblock in: \bibinfo{booktitle}{Proceedings of the 2013 conference on
  empirical methods in natural language processing}, \bibinfo{year}{2013}, pp.
  \bibinfo{pages}{1533--1544}.
\bibitem[{Wang et~al.(2015)Wang, Berant, and Liang}]{wang2015building}
\bibinfo{author}{Y.~Wang}, \bibinfo{author}{J.~Berant},
  \bibinfo{author}{P.~Liang},
\newblock \bibinfo{title}{{Building a semantic parser overnight}},
\newblock in: \bibinfo{booktitle}{Proceedings of the 53rd Annual Meeting of the
  Association for Computational Linguistics and the 7th International Joint
  Conference on Natural Language Processing (Volume 1: Long Papers)},
  \bibinfo{year}{2015}, pp. \bibinfo{pages}{1332--1342}.
\bibitem[{Su et~al.(2016)Su, Sun, Sadler, Srivatsa, Gur, Yan, and
  Yan}]{suGeneratingCharacteristicrichQuestion2016}
\bibinfo{author}{Y.~Su}, \bibinfo{author}{H.~Sun}, \bibinfo{author}{B.~Sadler},
  \bibinfo{author}{M.~Srivatsa}, \bibinfo{author}{I.~Gur},
  \bibinfo{author}{Z.~Yan}, \bibinfo{author}{X.~Yan},
\newblock \bibinfo{title}{On {{Generating Characteristic-rich Question Sets}}
  for {{QA Evaluation}}},
\newblock in: \bibinfo{booktitle}{Proceedings of the 2016 {{Conference}} on
  {{Empirical Methods}} in {{Natural}} {{Language Processing}}},
  \bibinfo{publisher}{{Association for Computational Linguistics}},
  \bibinfo{address}{{Austin, Texas}}, \bibinfo{year}{2016}, pp.
  \bibinfo{pages}{562--572}. \DOIprefix\doi{10.18653/v1/D16-1054}.
\bibitem[{Dubey et~al.(2019)Dubey, Banerjee, Abdelkawi, and
  Lehmann}]{dubeyLCQuADLargeDataset2019}
\bibinfo{author}{M.~Dubey}, \bibinfo{author}{D.~Banerjee},
  \bibinfo{author}{A.~Abdelkawi}, \bibinfo{author}{J.~Lehmann},
\newblock \bibinfo{title}{{{LC-QuAD}} 2.0: {{A Large Dataset}} for {{Complex
  Question Answering}} over {{Wikidata}} and {{DBpedia}}},
\newblock in: \bibinfo{editor}{C.~Ghidini}, \bibinfo{editor}{O.~Hartig},
  \bibinfo{editor}{M.~Maleshkova}, \bibinfo{editor}{V.~Sv{\'a}tek},
  \bibinfo{editor}{I.~Cruz}, \bibinfo{editor}{A.~Hogan},
  \bibinfo{editor}{J.~Song}, \bibinfo{editor}{M.~Lefran{\c c}ois},
  \bibinfo{editor}{F.~Gandon} (Eds.), \bibinfo{booktitle}{The {{Semantic Web}}
  \textendash{} {{ISWC}} 2019}, volume \bibinfo{volume}{11779},
  \bibinfo{publisher}{{Springer International Publishing}},
  \bibinfo{address}{{Cham}}, \bibinfo{year}{2019}, pp. \bibinfo{pages}{69--78}.
  \DOIprefix\doi{10.1007/978-3-030-30796-7_5}.
\bibitem[{Gu et~al.(2021)Gu, Kase, Vanni, Sadler, Liang, Yan, and
  Su}]{guThreeLevelsGeneralization2021}
\bibinfo{author}{Y.~Gu}, \bibinfo{author}{S.~Kase}, \bibinfo{author}{M.~Vanni},
  \bibinfo{author}{B.~Sadler}, \bibinfo{author}{P.~Liang},
  \bibinfo{author}{X.~Yan}, \bibinfo{author}{Y.~Su},
\newblock \bibinfo{title}{Beyond {{I}}.{{I}}.{{D}}.: {{Three Levels}} of
  {{Generalization}} for {{Question Answering}} on {{Knowledge Bases}}},
\newblock in: \bibinfo{booktitle}{Proceedings of the {{Web Conference}} 2021},
  \bibinfo{publisher}{{ACM}}, \bibinfo{address}{{Ljubljana Slovenia}},
  \bibinfo{year}{2021}, pp. \bibinfo{pages}{3477--3488}.
  \DOIprefix\doi{10.1145/3442381.3449992}.
\bibitem[{Cao et~al.(2022)Cao, Shi, Pan, Nie, Xiang, Hou, Li, He, and
  Zhang}]{caoKQAProDataset2022}
\bibinfo{author}{S.~Cao}, \bibinfo{author}{J.~Shi}, \bibinfo{author}{L.~Pan},
  \bibinfo{author}{L.~Nie}, \bibinfo{author}{Y.~Xiang},
  \bibinfo{author}{L.~Hou}, \bibinfo{author}{J.~Li}, \bibinfo{author}{B.~He},
  \bibinfo{author}{H.~Zhang},
\newblock \bibinfo{title}{{KQA} pro: {A} dataset with explicit compositional
  programs for complex question answering over knowledge base},
\newblock in: \bibinfo{booktitle}{Proceedings of the 60th Annual Meeting of the
  Association for Computational Linguistics}, \bibinfo{publisher}{Association
  for Computational Linguistics}, \bibinfo{address}{Dublin, Ireland},
  \bibinfo{year}{2022}, pp. \bibinfo{pages}{6101--6119}.
  \DOIprefix\doi{10.18653/v1/2022.Association for Computational
  Linguistics-long.422}.
\bibitem[{Keysers et~al.(2020)Keysers, Sch{\"a}rli, Scales, Buisman, Furrer,
  Kashubin, Momchev, Sinopalnikov, Stafiniak, Tihon, Tsarkov, Wang, {van Zee},
  and Bousquet}]{keysersMeasuringCompositionalGeneralization2020}
\bibinfo{author}{D.~Keysers}, \bibinfo{author}{N.~Sch{\"a}rli},
  \bibinfo{author}{N.~Scales}, \bibinfo{author}{H.~Buisman},
  \bibinfo{author}{D.~Furrer}, \bibinfo{author}{S.~Kashubin},
  \bibinfo{author}{N.~Momchev}, \bibinfo{author}{D.~Sinopalnikov},
  \bibinfo{author}{L.~Stafiniak}, \bibinfo{author}{T.~Tihon},
  \bibinfo{author}{D.~Tsarkov}, \bibinfo{author}{X.~Wang},
  \bibinfo{author}{M.~{van Zee}}, \bibinfo{author}{O.~Bousquet},
  \bibinfo{title}{Measuring {{Compositional Generalization}}: {{A Comprehensive
  Method}} on {{Realistic Data}}}, \bibinfo{year}{2020}.
  \href{http://arxiv.org/abs/1912.09713}{{\tt arXiv:1912.09713}}.
\bibitem[{Ley(2002)}]{leyDBLPComputerScience2002}
\bibinfo{author}{M.~Ley},
\newblock \bibinfo{title}{The {{DBLP Computer Science Bibliography}}:
  {{Evolution}}, {{Research Issues}}, {{Perspectives}}},
\newblock in: \bibinfo{editor}{G.~Goos}, \bibinfo{editor}{J.~Hartmanis},
  \bibinfo{editor}{J.~{van Leeuwen}}, \bibinfo{editor}{A.~H.~F. Laender},
  \bibinfo{editor}{A.~L. Oliveira} (Eds.), \bibinfo{booktitle}{String
  {{Processing}} and {{Information Retrieval}}}, volume \bibinfo{volume}{2476},
  \bibinfo{publisher}{{Springer Berlin Heidelberg}}, \bibinfo{address}{{Berlin,
  Heidelberg}}, \bibinfo{year}{2002}, pp. \bibinfo{pages}{1--10}.
  \DOIprefix\doi{10.1007/3-540-45735-6_1}.
\bibitem[{Banerjee et~al.(2022)Banerjee, Nair, Kaur, Usbeck, and
  Biemann}]{banerjeeModernBaselinesSPARQL2022}
\bibinfo{author}{D.~Banerjee}, \bibinfo{author}{P.~A. Nair},
  \bibinfo{author}{J.~N. Kaur}, \bibinfo{author}{R.~Usbeck},
  \bibinfo{author}{C.~Biemann},
\newblock \bibinfo{title}{Modern {{Baselines}} for {{SPARQL Semantic
  Parsing}}},
\newblock in: \bibinfo{booktitle}{Proceedings of the 45th {{International ACM
  SIGIR Conference}} on {{Research}} and {{Development}} in {{Information
  Retrieval}}}, \bibinfo{year}{2022}, pp. \bibinfo{pages}{2260--2265}.
  \DOIprefix\doi{10.1145/3477495.3531841}.
  \href{http://arxiv.org/abs/2204.12793}{{\tt arXiv:2204.12793}}.
\bibitem[{Raffel et~al.(2020)Raffel, Shazeer, Roberts, Lee, Narang, Matena,
  Zhou, Li, and Liu}]{raffelExploringLimitsTransfer}
\bibinfo{author}{C.~Raffel}, \bibinfo{author}{N.~Shazeer},
  \bibinfo{author}{A.~Roberts}, \bibinfo{author}{K.~Lee},
  \bibinfo{author}{S.~Narang}, \bibinfo{author}{M.~Matena},
  \bibinfo{author}{Y.~Zhou}, \bibinfo{author}{W.~Li}, \bibinfo{author}{P.~J.
  Liu},
\newblock \bibinfo{title}{Exploring the {{Limits}} of {{Transfer Learning}}
  with a {{Unified Text-to-Text Transformer}}},
\newblock \bibinfo{journal}{J. Mach. Learn. Res.} \bibinfo{volume}{21}
  (\bibinfo{year}{2020}) \bibinfo{pages}{1--67}.

\end{thebibliography}

\appendix

\end{document}